# A Secure Variant of the Hill Cipher [†]

Mohsen Toorani [‡]

Abolfazl Falahati

**Abstract**

*The Hill cipher is a classical symmetric encryption algorithm that succumbs to the know-plaintext attack. Although its vulnerability to cryptanalysis has rendered it unusable in practice, it still serves an important pedagogical role in cryptology and linear algebra. In this paper, a variant of the Hill cipher is introduced that makes the Hill cipher secure while it retains the efficiency. The proposed scheme includes a ciphering core for which a cryptographic protocol is introduced.*

## 1. Introduction

The Hill cipher was invented by L.S. Hill in 1929 [1]. It is a famous polygram and a classical symmetric cipher based on matrix transformation but it succumbs to the known-plaintext attack [2]. Although its vulnerability to cryptanalysis has rendered it unusable in practice, it still serves an important pedagogical role in both cryptology and linear algebra. The Hill cipher is a block cipher that has several advantages such as disguising letter frequencies of the plaintext, its simplicity because of using matrix multiplication and inversion for encryption and decryption, and its high speed and high throughput [3].

Several researches have been done to improve the security of the Hill cipher. Yeh et al. [4] used two co-prime base numbers that are securely shared between the participants. Although their scheme thwarts the known-plaintext attack, it is so time-consuming, requires many mathematical manipulations, and is not efficient especially when dealing with a bulk of data. Saeednia [5] tried to make the Hill cipher secure using some random permutations of columns and rows of the key matrix but it is proved that his cryptosystem is vulnerable to the known-plaintext attack [6], the same vulnerability of the original Hill cipher. Lin et al. [6] tried to improve the security of the Hill cipher using several random numbers generated in a hash chain but their scheme is not efficient. Ismail et al. [3] used an initial vector that multiplies successively by some orders of the key matrix to produce the corresponding key of each block but it has several security problems [7].

In this paper, a secure cryptosystem is introduced that overcomes all the security drawbacks of the Hill cipher. The proposed scheme includes an encryption algorithm that is a variant of the Affine Hill cipher for which a secure cryptographic protocol is introduced. The encryption core of the proposed scheme has the same structure of the Affine Hill cipher but its internal manipulations are different from the previously proposed cryptosystems. The rest of this paper is organized as follows. Section 2 briefly introduces the Hill cipher. Our proposed scheme is introduced and its computational costs are evaluated in Section 3, and Section 4 concludes the paper.

## 2. The Hill Cipher

In the Hill cipher, the ciphertext is obtained from the plaintext by means of a linear transformation. The plaintext row vector $\mathbf{X}$ is encrypted as $\mathbf{Y} = \mathbf{XK} \pmod{m}$ in which $\mathbf{Y}$ is the ciphertext row vector, $\mathbf{K}$ is an $n \times n$ key matrix where $k_{ij} \in Z_m$ in which $Z_m$ is ring of integers modulo $m$ where $m$ is a natural number that is greater than one. The encryption procedure proceeds by encoding the resulted ciphertext row vector into alphabets of the main plaintext. The value of the modulus $m$ in the original Hill cipher was 26 but its value can be optionally selected. The key matrix $\mathbf{K}$ is supposed to be securely shared between the participants. The ciphertext $\mathbf{Y}$ is decrypted as $\mathbf{X} = \mathbf{YK}^{-1} \pmod{m}$. All operations are performed over $Z_m$.

For decryption to be possible, the key matrix $\mathbf{K}$ should be invertible or equivalently, it should satisfy $\gcd(\det \mathbf{K} \pmod{m}, m) = 1$ [2]. However, many of square matrices are not invertible over $Z_m$. The risk of determinant having common factors with the modulus can be reduced by taking a prime number as the modulus. Such selection also increases the keyspace of the cryptosystem [8].

The security of the Hill cipher depends on confidentiality of the key matrix **K** and its rank $n$. When $n$ is unknown and the modulus $m$ is not too large, the opponent could simply try successive values of $n$ until he founds the key. If the guessed value of $n$ was incorrect, the obtained key matrix would be disagreed with further plaintext-ciphertext pairs. The most important security flaw of the Hill cipher is regarded to its vulnerability to the known-plaintext attack. It can be broken by taking just $n$ distinct pairs of plaintext and ciphertext [2]. In this kind of attack, the cryptanalyst possesses the plaintext of some messages and the corresponding ciphertext of those messages. He will try to deduce the key or an algorithm to decrypt any new messages encrypted with the same key.

The *Affine Hill cipher* extends the concept of Hill cipher by mixing it with a nonlinear affine transformation [2] so the encryption expression will have the form of $\mathbf{Y} = \mathbf{XK} + \mathbf{V}(\bmod\, m)$. In this paper, we extend such concept to introduce a secure variant of the Hill cipher.

## 3. The Proposed Scheme

The proposed cryptosystem includes a ciphering core that is depicted in Figure 1, and a one-pass protocol which is shown in Figure 2. The encryption core has the same structure of the Affine Hill cipher but in order to give more randomization to the introduced scheme and to strengthen it against the common attacks, each block of data is encrypted using a random number. For avoiding multiple random number generations, only one random number is generated at the beginning of encryption and the corresponding random number of the following data blocks is recursively generated using a one-way hash function in a hash chain, as it is depicted in Figure 1. The basic random number that is generated prior to the encryption should be securely shared between the participants. This can be done using the introduced one-pass protocol that is depicted in Figure 2 where the encryption and decryption procedures should be followed from Figure 1. The steps will be:

1. Alice secretly selects random integers $a_0$ and $b$ where $0 < a_0 < p-1$ and $1 < b < n^2$ in which $n$ is the rank of the key matrix. She computes $r = a_0 k_{ij}(\bmod\, p)$ where $i = \lceil b/n \rceil$ and $j = b - n(i-1)$ in which $\lceil . \rceil$ denotes the ceiling. She encodes the plaintext message into some row vectors $\mathbf{X} = [x_1\ x_2\ ...\ x_n]$. For the $t^{th}$ block of data to be encrypted ($t = 1,2,...$), she computes $a_t$ with a recursive expression as

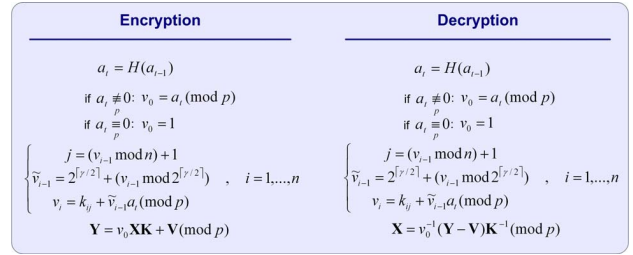

**Figure 1. Ciphering core of the proposed scheme**

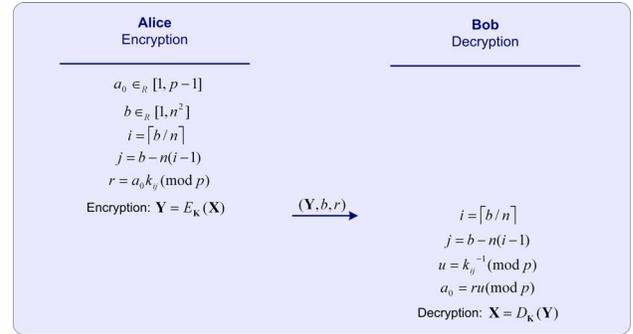

**Figure 2. A protocol for the proposed scheme**

$a_t = H(a_{t-1})$ in which $H(.)$ denotes the one-way hash function. If $a_t$ is invertible mod $p$, i.e. $a_t \not\equiv 0 (\bmod\, p)$, she puts $v_0 = a_t(\bmod\, p)$. Otherwise, she puts $v_0 = 1$. She produces the row vector $\mathbf{V} = [v_1\ v_2\ ...\ v_n]$ with the recursive expression as $v_i = k_{ij} + \tilde{v}_{i-1} a_t (\bmod\, p)$ for $i = 1,...,n$ and $j = (v_{i-1} \bmod n)+1$, in which $\tilde{v}_{i-1}$ is defined as $\tilde{v}_{i-1} = 2^{\lceil \gamma/2 \rceil} + (v_{i-1} \bmod 2^{\lceil \gamma/2 \rceil})$ where $\gamma = \lfloor \log_2 v_{i-1} \rfloor + 1$ denotes the bit-length of $v_{i-1}$ and $\lfloor . \rfloor$ indicates the floor. She then encrypts all the plaintext vectors as $\mathbf{Y} = v_0 \mathbf{XK} + \mathbf{V}(\bmod\, p)$. She repeats the procedure until all blocks of plaintext become encrypted.

2. Bob computes $u = k_{ij}^{-1}(\bmod\, p)$ and $a_0 = ru(\bmod\, p)$ in which $i = \lceil b/n \rceil$ and $j = b - n(i-1)$. He uses $a_0$ for decrypting the ciphertext as $\mathbf{X} = v_0^{-1}(\mathbf{Y} - \mathbf{V})\mathbf{K}^{-1}(\bmod\, p)$, as it is depicted in Figure 1.

The introduced protocol is a one-pass protocol that is designed for the proposed cryptosystem. As a one-pass protocol, it does not have any explicit authentication step but it is secure, and does not reveal any secret information. It is so suitable for situations where both of participants are not online.

The introduced expression for generating the elements of vector **V** as $v_i = k_{ij} + \tilde{v}_{i-1} a_t \pmod{p}$ and defining $\tilde{v}_{i-1}$ as $\tilde{v}_{i-1} = 2^{\lceil \gamma/2 \rceil} + (v_{i-1} \bmod 2^{\lceil \gamma/2 \rceil})$ takes advantages of ideas behind the MQV key-exchange protocols [9]. $\tilde{v}_{i-1}$ is simply computed by taking the least significant half in binary representation of $v_{i-1}$ and such definition will decrease the computational costs and consequently, increases the efficiency [9].

The proposed cryptosystem neutralizes all the security drawbacks of the Hill cipher. It thwarts the known-plaintext attack since *n* equations cannot be used for solving an unknown $n \times n$ matrix and *2n* unknown parameters. Choosing a large prime number *p* as the modulus extremely enhances the keyspace so the brute-force or equivalently, the ciphertext-only attack does not have any benefit for the attacker. The random number after a secure transmission is recursively encoded with the one-way hash function so it differs for each block of plaintext. The chosen-ciphertext and chosen-plaintext attacks are also thwarted since the random number $a_0$ that its knowledge is essential for such attacks, is exchanged through a secure protocol.

### 3.1. Computational Costs

Let $T_{Enc}$ and $T_{Dec}$ denote the running time for encryption and decryption of each block of data respectively. By neglecting the computational costs of the introduced protocol and the required computations of computing the inverse key matrix that is used for decryption, we have:

$$T_{Enc} \cong (n^2 + 2n)T_{Mul} + (n^2 + n + 1)T_{Add} + T_{Hash} \quad (1)$$

$$T_{Dec} \cong (n^2 + 2n)T_{Mul} + (n^2 + n + 1)T_{Add} + T_{Hash} + T_{Inv} \quad (2)$$

in which $T_{Hash}$ is the running time for the hash calculations, and $T_{Mul}$, $T_{Add}$ and $T_{Inv}$ are the time needed for the scalar modular multiplication, addition, and inverse calculations respectively. Total required number of operations for computing SHA-1 and MD5 hash functions are calculated as 1110 and 744 operations respectively [10]. Each of $T_{Mul}$, $T_{Add}$ and $T_{Inv}$ requires different number of operations. Let $\zeta = \lfloor \log_2 p \rfloor + 1$ denotes the bit-length of modulus *p*. Using the conventional methods, the running time for calculating a modular addition, modular multiplication, and modular inverse will be of [11]:

$$T_{Add} = O(\zeta) \quad (3)$$

$$T_{Mul} = O(\zeta^2) \quad (4)$$

$$T_{Inv} = O(\zeta^3) \quad (5)$$

There are also many fast algorithms for the computations [12] but we consider the time complexity of conventional methods since it corresponds with the worst situation and anyone can decrease the required number of operations by deploying faster algorithms. The computational complexity of the proposed scheme for encrypting and decrypting each block of data can be simply estimated by substituting expressions (3-5) into (1) and (2). The running time for encryption and decryption of each block of data explicitly depends on $\zeta$ (the bit-length of modulus *p*) and *n* (the rank of the key matrix). The size of modulus *p* depends on the number of deployed alphabets in the plaintext.

Let *L* denotes the number of letters in the plaintext. Total processing time for enciphering the whole blocks of plaintext is:

$$T_{Total\_Enc} \cong \left\lceil \frac{L}{n} \right\rceil \left( (n^2 + 2n)T_{Mul} + (n^2 + n + 1)T_{Add} + T_{Hash} \right) \quad (6)$$

while the running time for decrypting the whole ciphertext is:

$$T_{Total\_Dec} \cong \left\lceil \frac{L}{n} \right\rceil \left( (n^2 + 2n)T_{Mul} + (n^2 + n + 1)T_{Add} + T_{Hash} + T_{Inv} \right) \quad (7)$$

The computational costs of the proposed scheme for encryption/decryption of all blocks of data can be simply calculated by substituting expressions (3-5) into (6) and (7). Figure 3 depicts the effects of rank value of the key matrix on the total number of operations for encipherment /decipherment of the whole plaintext/ciphertext that is obtained using (6) and (7) for $L = 1000$ and $p = 257$. The size effects of the modulus *p* on the total number of operations for encipherment/decipherment of the whole plaintext/ciphertext is also depicted in Figure 4 that is obtained using (6) and (7) for $L = 1000$ and $n = 4$. It is noteworthy that the waves in Figure 3 are according to the introduced ceiling function in (6) and (7) while the steps in Figure 4 are due to logarithmic relationship between the modulus *p* and its bit-length $\zeta$. Table 1 gives a comparison between the required number of operations for encrypting/decrypting each block of data in the proposed scheme and those of the other schemes. It shows that the computational cost of the proposed scheme is slightly more than that of the Affine Hill cipher so it is computationally efficient while it thwarts the security vulnerabilities of the Hill cipher.

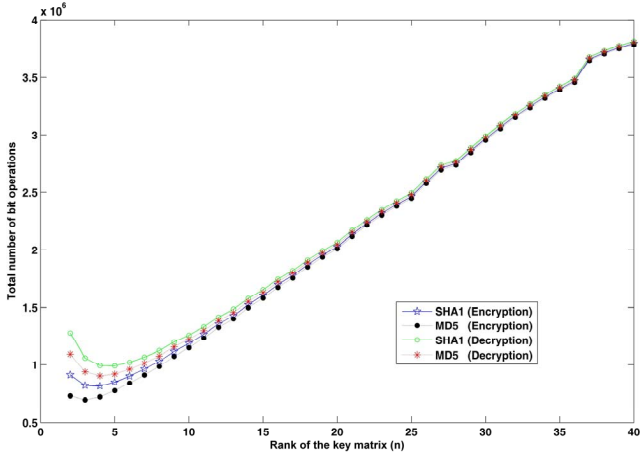

**Figure 3. Total number of operations required for encrypting a plaintext of *L*=1000 letters for different rank values and a fixed modulus (*p*=257).**

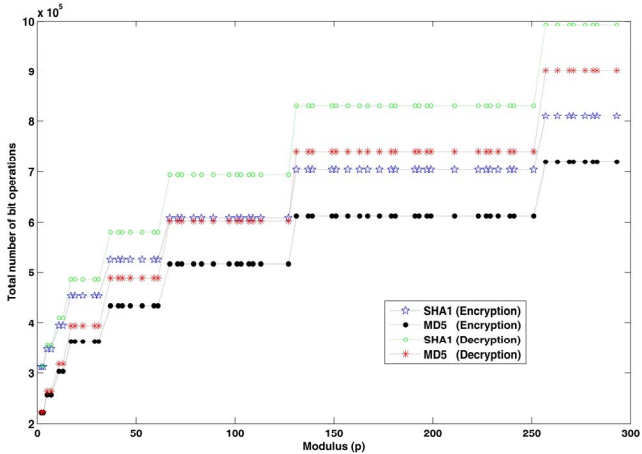

**Figure 4. Total number of operations required for encrypting a plaintext of *L*=1000 letters for different modulo *p* and a fixed rank value (*n*=4).**

**Table 1. Computational costs of different schemes for encryption/decryption of each block of data**

| Different Schemes | Operation | $T_{Mul}$ | $T_{Add}$ | $T_{Inv}$ | $T_{Hash}$ |
|---|---|---|---|---|---|
| Original Hill Cipher | Encryption | $n^2$ | $n^2 - n$ | - | - |
| | Decryption | $n^2$ | $n^2 - n$ | - | - |
| Affine Hill Cipher | Encryption | $n^2$ | $n^2$ | - | - |
| | Decryption | $n^2$ | $n^2$ | - | - |
| Lin et al.'s Scheme [6] | Encryption | $n^2 + n + 3$ | $n^2 + 4$ | - | $n+1$ |
| | Decryption | $n^2 + n + 3$ | $n^2 + 4$ | 1 | $n+1$ |
| The Proposed Scheme | Encryption | $n^2 + 2n$ | $n^2 + n + 1$ | - | 1 |
| | Decryption | $n^2 + 2n$ | $n^2 + n + 1$ | 1 | 1 |

## 4. Conclusions

In this paper, a symmetric cryptosystem is introduced that is actually a secure variant of the Affine Hill cipher. It includes a ciphering core for which a one-pass cryptographic protocol is introduced. The outer structure of the ciphering core is similar to the Affine Hill cipher but its inner manipulations are different. Each block of data is encrypted using a different random number that is generated using a chained hash function. The proposed cryptosystem thwarts the known-plaintext, chosen-ciphertext, and chosen-plaintext attacks. Since the modulus is a prime number, the keyspace is greatly increased and the ciphertext-only attack is also thwarted.